# Production Cross sections of $^{190-193}$Au Radioisotopes Produced From $^{11}$B+ $^{nat}$W Reactions Upto 63 MeV Projectile Energy

Dibyasree Choudhury[1], Susanta Lahiri[1,2*]

[1]Saha Institute of Nuclear Physics, 1/AF Bidhannagar, Kolkata-700064, India
[2]Homi Bhaba National Institute, 1/AF Bidhannagar, Kolkata-700064, India
*susanta.lahiri@saha.ac.in

**Abstract**

This paper reports a new route for production of $^{190-193}$Au radionuclides from $^{nat}$W target. For the first time the production cross section of $^{11}$B induced reactions on tungsten for production of $^{190-193}$Au were measured up to 63 MeV boron energy. High cross section was observed for $^{190,192}$Au at the highest energy where as $^{191,193}$Au showed moderate cross section values. The experimentally obtained cross sections were compared with the predictions of the PACE4 and EMPIRE3.2.2 code.

## 1. Introduction

In recent times, interest towards exploring auger emitters in medical science has gained momentum. Auger electrons having small range offers cytotoxicity only when incorporated in the cell's DNA or in the immediate vicinity of the nucleus of a cell. The low energy auger electrons allow targeted therapy in DNA level. $^{198}$Au mainly produced from reactors has been vastly explored as a therapeutic radionuclide [1] but on the other hand, the much less explored neutron deficient radionuclides of Au also offer suitable nuclear characteristics such as half lives, decay modes, high intensity auger electrons for their clinical applications.

Earlier lighter projectiles such as $^{1}H$, $^{2}H$ or $^{4}He$ particles were used for the production of $^{190-193}Au$ radioisotopes. Takács et al [2] produced $^{191-196}Au$ radioisotopes from α-particle irradiated iridium target and studied the excitation function of (α,xn) reaction upto 50 MeV projectile energy. However, production of Au radionuclides from Ir was studied along with several co-produced radioisotopes and the paper mainly focussed on the production of $^{195m}Pt$. Ditroi et al [3] produced $^{196m,196g(cum),195g(cum),194,191(cum)}Au$ from Au target by (p,pxn) reaction using proton beam upto 65 MeV. They further compared the experimental results with theoretical codes TALYS1.6 and EMPIRE 3.2 code. The Au radioisotopes produced were not no-carrier-added (NCA), thereby limiting their further application. Tarkanyi et al [4] reported the production of $^{191-196,198}Au$ from $^{nat}Pt$ target using $^{1}H$ beam upto 70 MeV. Maximum cross section of 530 mb at 42.4 MeV, 673 mb at 36.7 MeV, 454 mb at 22.6 MeV for $^{191}Au$, $^{192}Au$ and $^{193}Au$ were reported respectively. In the same year, this group also explored an alternative production route of $^{192-196}Au$ by deuteron bombardment on $^{nat}Pt$ target upto 21 MeV projectile energy [5]. It was found that the cross section was higher for proton induced reactions than deuteron induced ones. But both these papers reported the production of Au along with various other radioisotopes. Indirect production routes of $^{190-193}Au$ through spallation reaction on thick cylindrical Pb target by 660 MeV protons ($10^{13}$ protons/s) or as a decay product of Hg isotopes were also reported by Majerle et al. [6]. Several radionuclides were also co-produced along with $^{190-193}Au$ radionuclides.

For last two decades our laboratory made pioneering contribution on developing the neutron-deficient isotope production route by heavy ion activation. A concise literature can be found in reference [7]. For example, in the context of gold radionuclide production, our laboratory reported production of $^{192,193}Au$ as daughter product of $^{192,193}Hg$ which was produced by irradiating tantalum target with 95 MeV $^{16}O$ beam. However, they did not measure the corresponding excitation function [8]. In the present work, an attempt has been made to systematically measure the production cross section of $^{nat}W(^{11}B,xn)$ reaction upto 63 MeV projectile energy and investigate an alternate production route of the $^{190\text{-}193}Au$ radionuclides. This is the first report on use of boron beam to produce NCA Au radioisotopes. Interestingly it was found that this production route selectively produced only Au radioisotopes without any radioisotope of other elements.

## 2. Experimental

2.1 **Irradiation parameters:** Five $WO_3$ targets (936 μg/cm$^2$) were prepared by its deposition on Al foil (7.3 mg/cm$^2$) via evaporation technique at Tata Institute of Fundamental Research (TIFR) target laboratory. The targets were irradiated one by one with 42-63 MeV $^{11}B$ beam at BARC-TIFR Pelletron, Mumbai, India. The conventional stack foil technique was not used in this experiment to avoid any energy spread [9]. The exit energy of the projectile was calculated by the software SRIM [10]. The irradiation details are mentioned in Table 1. The incident beam was well-collimated on the target holder and the total charge of the incident particle was measured with the help of Faraday cup placed at the rear end of the target in conjunction with a current integrator (Danfysik). Series of γ- spectra were taken in a p-type HPGe detector of 2.35 keV resolution at 1.33 MeV. The sample to detector distance was ~10 cm. The radionuclides were identified from their corresponding photo peaks and decay data. The energy and efficiency calibration of the detector were performed using $^{152}Eu$ ($T_{1/2}$=13.3 a) standard source of known activity.

**2.2 Measurement of experimental cross section:** The production cross sections of $^{nat}W(^{11}B,xn)^{190\text{-}193}Au$ reaction was measured as a function of $^{11}B$ incident energy by using the following activation equation:

$$A = n\sigma(E)I(1 - e^{-\lambda T})$$

Where,
A= Activity (Bq) of a particular radionuclide at EOB
σ(E)= cross section of production of the radionuclide at incident energy E
I= Intensity of boron beam in particles/s
n= no. of atoms/cm$^2$

$\lambda$= disintegration constant in $s^{-1}$
T= Duration of irradiation in s

**2.3 Error calculation:** The uncertainty of each cross section was estimated by considering errors from all sources. Error due to efficiency calibration was insignificant (~0.5%) and hardly has any influence on the total uncertainty. Error due to counting statistics was different for different isotopes at different energies, $^{190}$Au (1.42-9.14%), $^{191}$Au (6.95-27.4%), $^{192}$Au (0.54-11.5%), $^{193}$Au (18.31-64.5%). Error in determining the target thickness was ~5%. The combined uncertainties due to beam current, incident beam energy, etc. were ~10%.

**2.4 Theoretical cross section calculation:** The production cross section of $^{190-193}$Au was compared with theoretical Monte Carlo codes PACE4 [11] and EMPIRE3.2.2 [12]. The evaporation code EMPIRE 3.2.2 (version Malta) was designed for simulation of the photons, nucleons, deuterons, tritons, helions, alpha, and light or heavy ions in the low and intermediate energy range. This code employs various nuclear reaction models such as optical model, coupled channels, etc., assuming that all reactions proceed via formation of compound nucleus. In the EMPIRE code, exciton model with 1.5 mean free path and EGSM (Enhanced Generalized Superfluid Model) level density were used.

The fusion-evaporation code PACE4, a modified version of PACE (projection angular momentum coupled evaporation) was used to calculate the excitation function of residues expected to be produced in $^{11}$B-induced reactions on a $^{nat}$W target. Fission is considered as a decay mode in this Monte Carlo statistical model code. The finite-range fission barrier of Sierk has been used. Total 20000 cascades were selected for each simulation. The code internally decides the level densities and masses it needs during de-excitation. The level density parameter was kept A/10 (A is mass number and 10 is free adjustable parameter) in this calculation.

## 3. Results and Discussions

$^{nat}W$ has five stable isotopes, $^{180}W$(0.12%), $^{182}W$ (26.50%), $^{183}W$ (14.31%). $^{184}W$(30.64%), $^{186}W$(28.43%). Therefore while calculating theoretical production cross sections, weighted average of each isotope was considered. Time resolved gamma-ray spectroscopy of $^{11}B$ irradiated $^{nat}WO_3$ target indicated the presence of $^{190-193}Au$ radioisotopes (Figure 1). The nuclear characteristics of each of the produced radioisotope have been shown in Table 2. Absence of characteristic photopeaks of radioisotope of any other elements makes this production route specific for the Au radioisotopes. The characteristic gamma lines were used in the evaluation of yield at EOB for each investigated radionuclides. The gamma lines which were shared by two Au radionuclides were not included in calculations, even if have high intensity. For example: 316.5 keV has contribution of both $^{192}Au$ and $^{191}Au$, hence not taken into consideration.

Attempts were made to identify the production pathways mentioned in Table 2 responsible for the each isotope from PACE4 code. The cross section of each gold radioisotope from the different stable isotopes of W has been shown in Figure 2 (a-e). From the plots, it was possible to narrow down the most feasible pathway for each of the products. The following conclusions could be made from the plots:

(i) None of the isotopes would be produced from $^{180}W$.

(ii) Au-191 would be produced from $^{183}W(^{11}B,3n)$ or $^{184}W(^{11}B,4n)$ reaction while Au-192 would be produced from $^{184}W(^{11}B,3n)$ or $^{186}W(^{11}B,5n)$ reactions.

(iii) The contributing reaction for production of Au-193 is $^{186}W(^{11}B,4n)$.

### *3.1 Cross sections of $^{nat}W(^{11}B,xn)$ reaction*

The measured experimental cross section data is shown in Figure 3(a-d), together with the results of theoretical calculations. The radioisotopes of $^{190-193}Au$ may be produced by

$^{nat}W(^{11}B,xn)$ reaction via different reaction channels. The maximum production cross sections for $^{190\text{-}193}Au$ was different for each isotope. Since the irradiation time for targets 1, 3 and 5 are more than 5 half lives of $^{190}Au$, therefore in the cross section calculation saturation activity and integrated current at 228 min (5 half lives of $^{190}Au$) was considered at the time of cross section calculation. The cross section of Au-190 increased with increase of projectile energy and was maximum at 63 MeV incident energy (216 mb). The experimental cross section was comparable with both PACE and EMPIRE codes except at the highest energy where EMPIRE value was 1.3 times lower than the experimental value. The cross section of Au-191 was zero upto 42 MeV but increased from 48 MeV upto 63 MeV with increasing energy. The highest cross section of 76 mb was observed at 63 MeV. PACE4 and EMPIRE3.2.2 over predicted the experimental results at 52 and 57 MeV. Au-192 followed similar trend as Au-190,191 with increase of cross section with energy. The highest cross section observed at 63 MeV was 201 mb. The theoretical data very much reproduced the experimental measurements. In case of Au-193, the complete excitation plot could be obtained at the experimental energy range studied. Highest cross section of 84 mb was obtained was obtained at 57 MeV. Cross sections predicted by the theoretical codes are matching with the experimental production cross section of Au-193. In fact, the theoretical cross section and the experimental matched for all the four radionuclides of Au except few points (e.g. 52 and 57 MeV for Au-191). This mismatch might be due to the experimental error.

**Conclusion**

Use of positron emitters in PET and alpha emitters in therapy is well established. It would also be of significance to exploit the tremendous potential of auger electrons for short range targeted therapy. This report offers a feasible production pathway for the no-carrier-added $^{190\text{-}193}Au$ radioisotopes which are high intensity auger emitters. Moderate production yield has been observed via this production pathway.

**Acknowledgement**

Authors gratefully acknowledge the support from TIFR target laboratory and staffs of BARC-TIFR Pelletron facility. This work is financially supported by the SINP-DAE 12 Five year plan: Trace, Ultratrace Analysis and Isotope Production (TULIP), Government of India.

**Table1.** Irradiation details

| Target+ ($WO_3$) | Projectile ($^{11}B$) energy, MeV | | | Irradiation time (h) | Integrated charge (µC) |
|---|---|---|---|---|---|
| | Incident Energy | Exit Energy | Energy at the centre of mass | | |
| 1 | 63 | 52.2 | 57.6 | 5.9 | 296 |
| 2 | 57 | 45.4 | 51.2 | 3.4 | 337 |
| 3 | 52 | 39.5 | 45.7 | 5.9 | 296 |
| 4 | 48 | 34.8 | 41.3 | 3.3 | 152 |
| 5 | 42 | 27.4 | 34.7 | 3.8 | 123 |

+target thickness was same for all, i.e. 936 µg/cm$^2$

**Table 2.** Decay characteristics of the produced radionuclides (https://www.nndc.bnl.gov/chart/chartNuc.jsp)

| Radionuclide | $T_{1/2}$ | Decay mode | $E_\gamma$,keV, (Intensity, %) | $E_{auger}$, keV (Intensity, %) | Probable production pathways | $E_{Threshold}$ (MeV) |
|---|---|---|---|---|---|---|
| Au-190 | 42.8 m | ε(100) | 295.7 (71)<br>301.8(23.4) | 7.24 (55.6)<br>51.0 (3.3) | $^{182}W(^{11}B,3n)$<br>$^{183}W(^{11}B,4n)$<br>$^{184}W(^{11}B,5n)$ | 32.8<br>39.3<br>47.2 |
| Au-191 | 3.18 h | ε(100) | 399.8(4.7%)<br>421.4(3.4%) | 7.24(86)<br>51.0 (4.4) | $^{183}W(^{11}B,3n)$<br>$^{184}W(^{11}B,4n)$ | 29.8<br>37.6 |
| Au-192 | 4.94 h | ε(100) | 316.5 (58%) | 7.24(56)<br>51.0 (3.4) | $^{184}W(^{11}B,3n)$<br>$^{186}W(^{11}B,5n)$ | 30.1<br>43.8 |
| Au-193 | 17.65 h | ε(100) | 268.2(3.6%) | 7.24 (78)<br>51.0 (4.5) | $^{186}W(^{11}B,4n)$ | 34.6 |

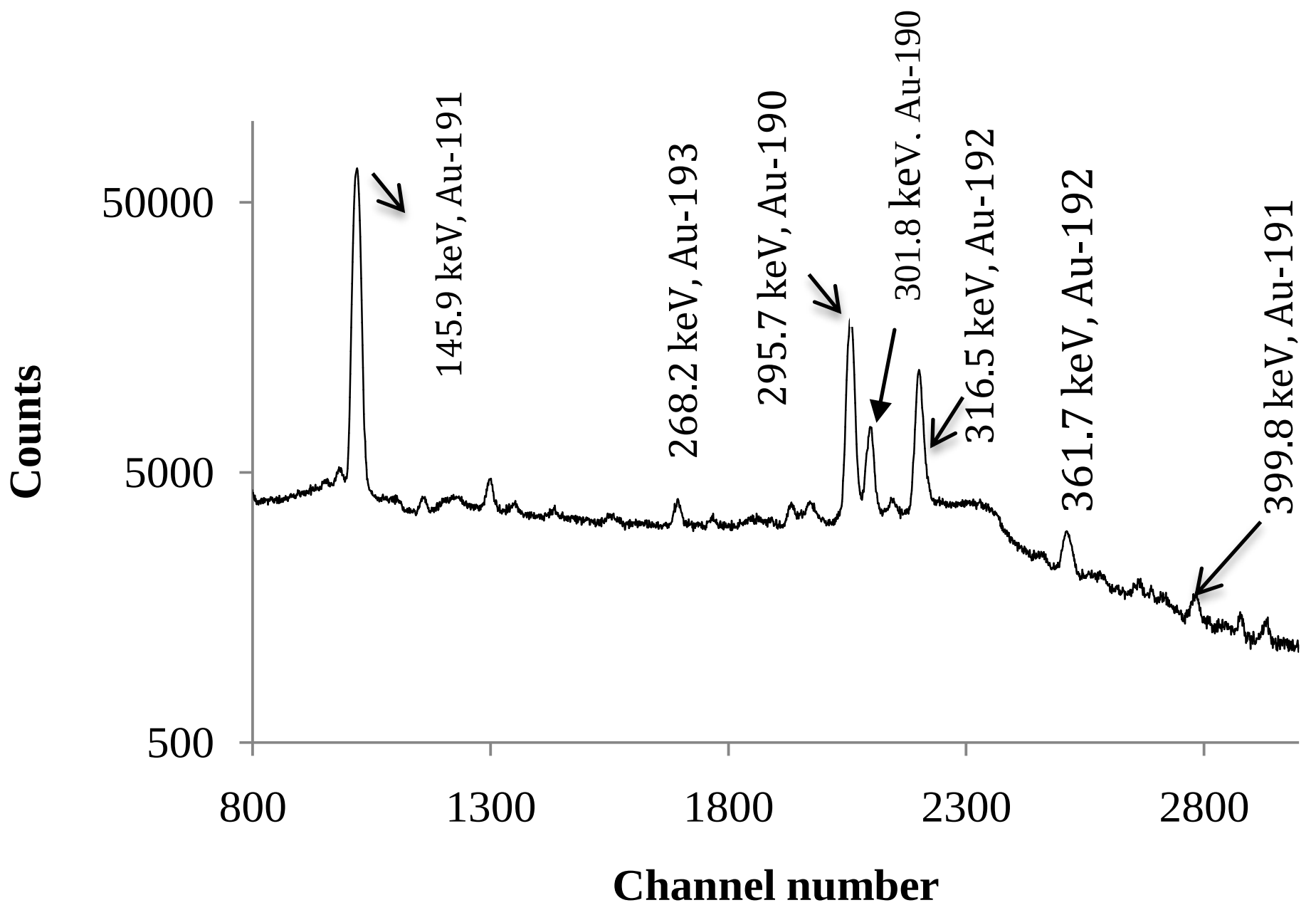

**Fig. 1.** Gamma spectrum of $^{11}B$ irradiated $^{nat}W$ target at 63 MeV incident projectile energy 0.5 h after EOB

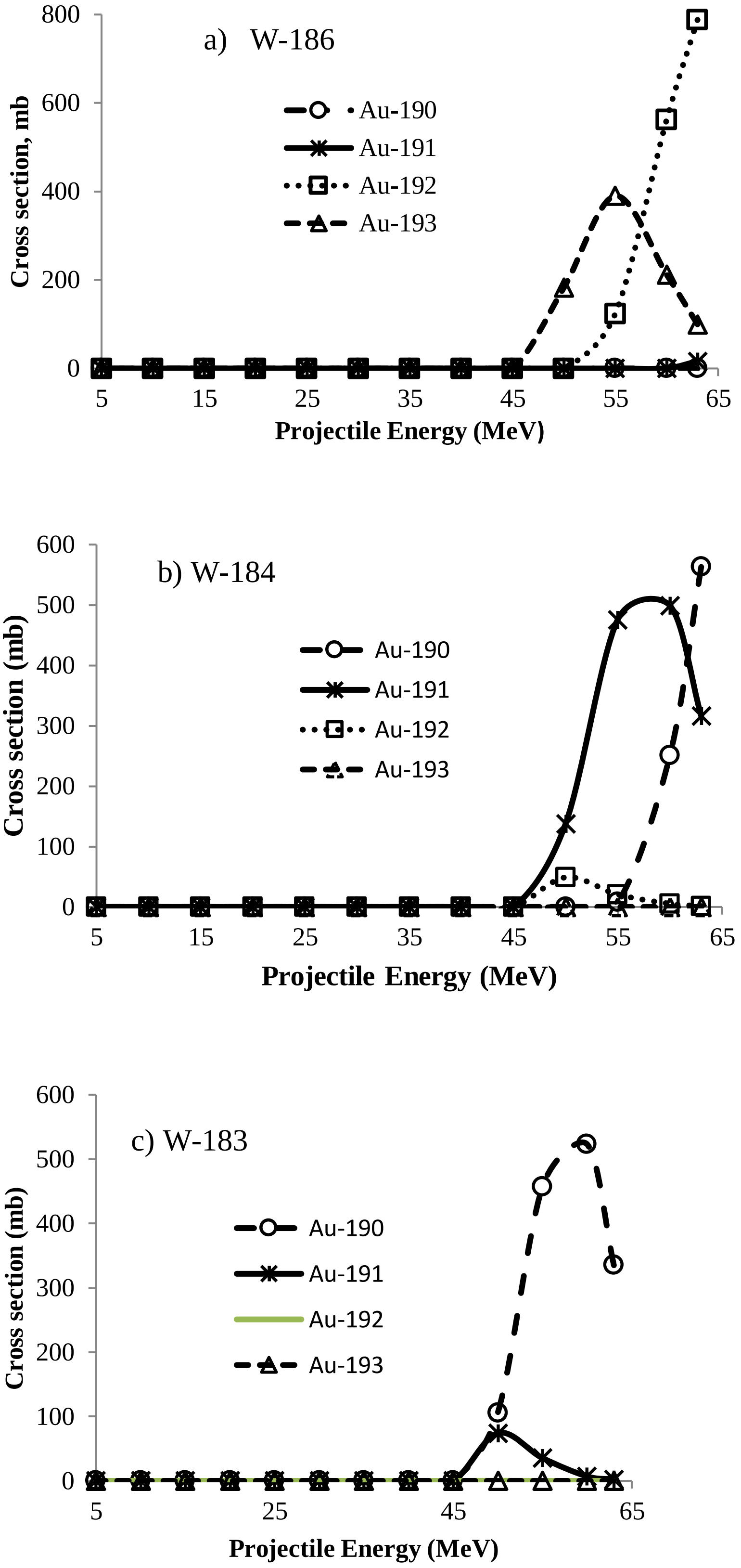

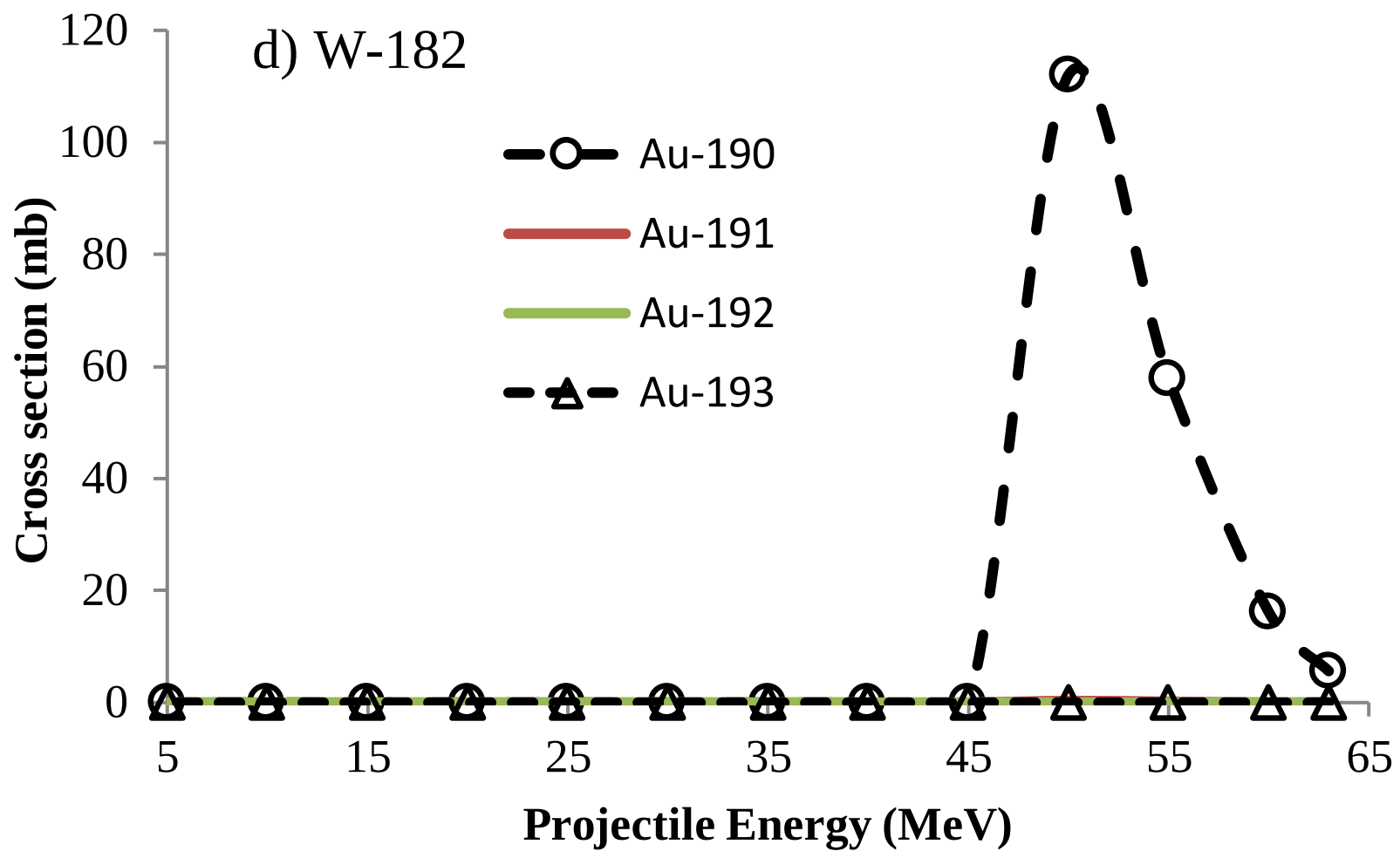

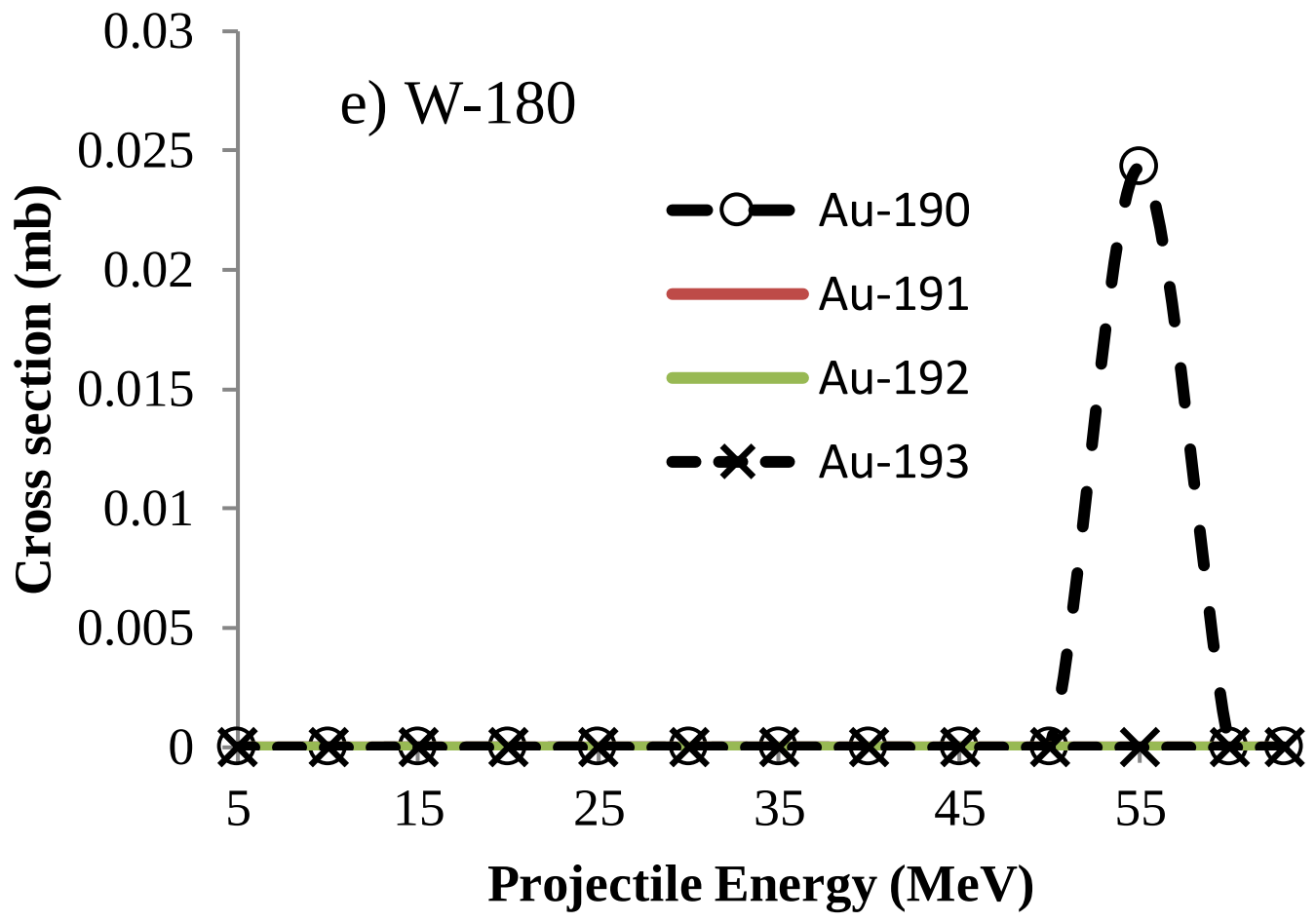

**Fig. 2.** Production cross section of $^{190\text{-}193}$Au from $^{186}$W (a), $^{184}$W (b), $^{183}$W (c), $^{182}$W (d), $^{180}$W (e) predicted by PACE 4

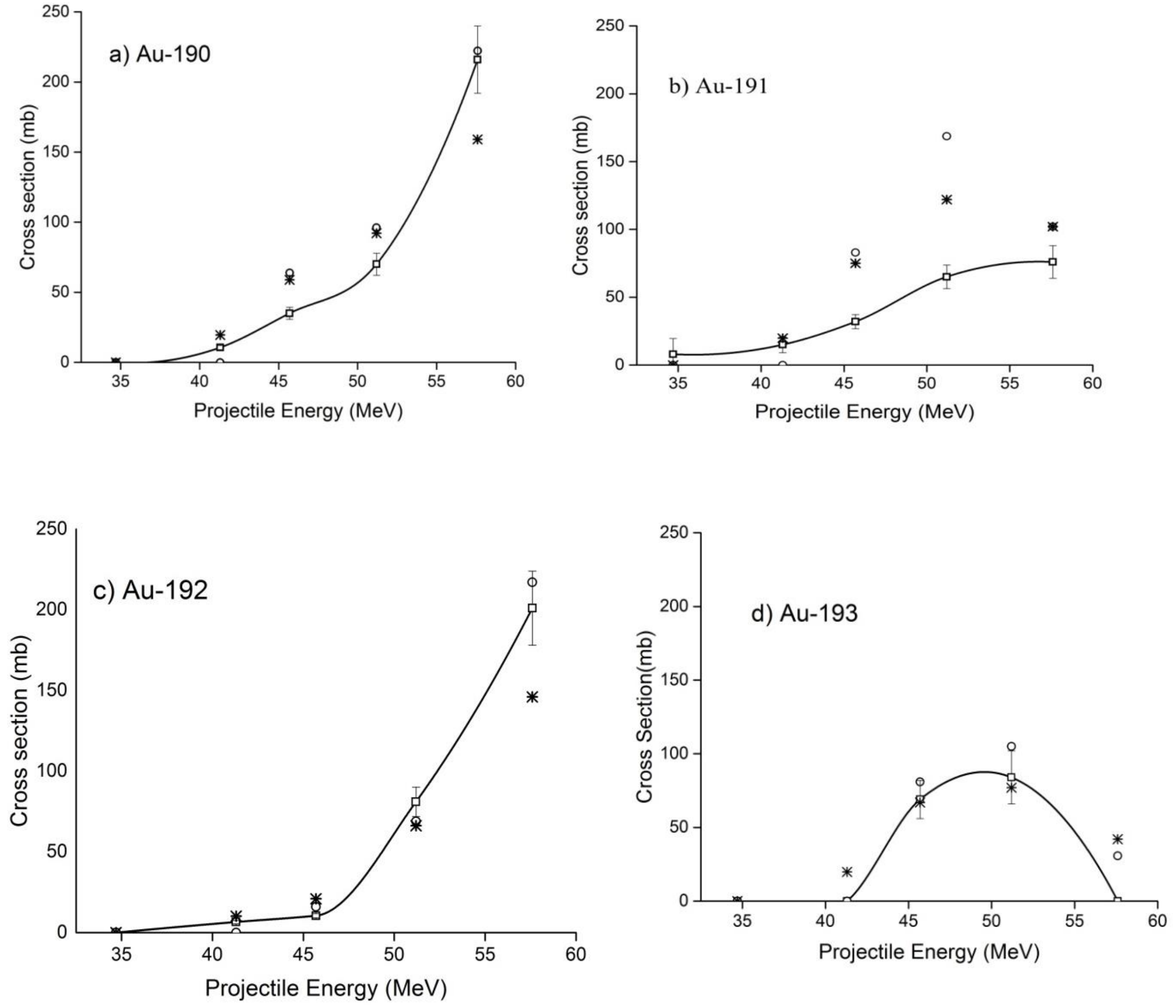

**Fig. 3(a-d).** Excitation function of $^{nat}W(^{11}B,xn)$ reaction from 42-63 MeV

—□— Experimental, o PACE4, * EMPIRE